\documentclass[preprint2]{aastex}

\def\Msun{\ifmmode{~{\rm M}_\odot}\else${\rm M}_\odot$~\fi}

\newcommand{\doo}[2]{{\frac{\partial#1}{\partial#2}}}
\newcommand{\fat}[1]{{\mathbf{{#1}}}}

\newcommand{\half}{{\scriptstyle{\frac12}}}

\newcommand{\LH}{{\Lambda}}

\shorttitle{Algorithmic Regularization with PN Terms}
\shortauthors{Mikkola and Merritt}

\begin{document}

\title{Implementing Few-Body Algorithmic Regularization
with Post-Newtonian Terms}

\author{Seppo Mikkola}
\affil{Tuorla Observatory, University of Turku, 
V\"ais\"al\"antie 20, Piikki\"o, Finland}
\email{Seppo.Mikkola@utu.fi}

\author{David Merritt}
\affil{Center for Computational Relativity and Gravitation
and Department of Physics,
Rochester Institute of Technology, Rochester, NY 14623, USA}
\email{David.Merritt@rit.edu}

\begin{abstract}
We discuss the implementation of a new regular algorithm
for simulation of the gravitational few-body problem.
The algorithm uses components from earlier methods,
including the chain structure, the logarithmic Hamiltonian, 
and the time-transformed leapfrog. 
This algorithmic regularization code, AR-CHAIN, 
can be used for the normal $N$-body problem,
as well as for problems with softened potentials and/or with 
velocity-dependent external perturbations, including
post-Newtonian terms, which we include up to order PN2.5.
Arbitrarily extreme mass ratios are allowed.
Only linear coordinate transformations are used and thus the algorithm is 
somewhat simpler than many earlier regularized schemes.
We present the results of performance tests which suggest that
the new code is either comparable in performance or superior to the 
existing regularization schemes based on the Kustaanheimo-Stiefel (KS) 
transformation.
This is true even for the two-body problem, independent of 
eccentricity. An important advantage of the new method 
is that, contrary to the older KS-CHAIN code, zero masses are allowed.
We use our algorithm to integrate the orbits of the S stars
around the Milky Way supermassive black hole for one million
years, including PN2.5 terms and an intermediate-mass black hole.
The three S stars with shortest periods are observed to
escape from the system after a few hundred thousand years.
\end{abstract}


\section{Introduction}

After the introduction of electronic computers, those who
carried out simulations of the gravitational $N$-body
problem soon realized that the classical methods of numerical
integration were often not satisfactorily accurate due simply to
the strong $1/r^2$ character of the gravitational force.

The situation changed when \citet{KS} published their (KS-) transformation
from four-dimensional to three-dimensional space.
The special case of planar system had been known for a long time
\citep{Levi-Civita} but it had turned out that a similar transformation
in the three-dimensional space was not possible.
The situation for the general $N$-body problem improved only after
the KS transformation became  well known, due in large part to 
publication by \citet{LinReg} of their text, which
comprehensively discussed the  application of the KS-transformation
to the perturbed two-body problem.
Further on, \citet{AZ}, \citet{Heggie}, \citet{Zare} and \cite{CHAIN} 
applied the KS transformation to the general three-body and $N$-body problems.

An entirely new way of regularizing close encounters
was invented simultaneously by
\citet{LogH,Algoreg} and by \citet{PretoTremaine1999}.
This new method introduced the so-called 
logarithmic Hamiltonian (LogH). 
Together with the simple leapfrog algorithm, 
this new method gives regular results for 
close encounters; in fact a correct trajectory is obtained 
for the two-body problem. 
Other, similar methods are described by \citet{HL97}. 

The remaining problem was that none of these regularization methods 
could be easily applied to systems with extremely large mass ratios.
In an attempt to solve this problem, \citet{TTL} introduced
the time-transformed leapfrog (TTL). This method is in some cases
mathematically equivalent to the LogH method, but it is more general,
and arbitrary mass ratios are allowed. The drawback of this
method, however, is that in some cases the roundoff error affects
the results considerably. In these new methods (LogH, TTL) the
regularization is achieved by using the leapfrog, hence the 
name ``algorithmic regularization.'' More details about the
thus-far mentioned methods can be found in the book by \citet{aa03}.

Since the leapfrog alone is rarely accurate enough, one must 
supplement the method with the extrapolation algorithm 
\citep{ Gragg1964, Gragg1965,BS,Press}, or with higher-order
leapfrogs \citep{Yoshida}, in order to get highly 
accurate results. 
Efficiency of the extrapolation procedure requires
that the basic algorithm (leapfrog in algorithmic regularization,
modified midpoint method in the KS-regularized codes) have
a certain symmetry. In the case of the leapfrog this means 
time reversibility. If the system has velocity-dependent 
forces, such as relativistic post-Newtonian (PN) terms 
\citep{Soffel1989},
then the required symmetry is more difficult to obtain.
One way to cure this problem was recently obtained
by \citet{GAR}, who formally doubled the dimensionality 
of the parameter space and constructed a generalized
midpoint method (GAR). This algorithm allows the 
use of algorithmic regularization and velocity-dependent
perturbations. However, these authors discussed the new algorithm
essentially only for the case of the perturbed two-body problem.

In this paper we discuss the details of our most recent
implementation of the algorithmic regularization method.
This uses the chain structure, the same structure as in the 
{\small KS-CHAIN} algorithm of \citet{CHAIN}. That device, together with
a new time-transformation function, significantly reduces
the roundoff problems and makes the new code a good alternative
for simulations of strongly-interacting few-body systems.
In the final section, we present some applications of
the chain routine to the problem of relativistic orbits around the
supermassive black hole at the Galactic center.

\section{Notation}
In this paper, we use the following basic symbols:
\[
\begin{tabular}{lll}
$G=1$      & gravitational constant\\
$t$        & time\\
$m_k$      & mass of body k \\
$\fat p_k$ &  momentum  \\
$\fat r_k$ &  position \\
$\fat v_k=\dot {\fat r_k}$ & velocity\\
$T=\sum_{k=1}^N{\fat p_k^2}/{2m_k}$ & kinetic energy\\ 
$U=\sum_{0<i<j\le N}{m_im_j}/{|\fat r_i-\fat r_j|} $ & potential energy\\
$E=T-U$ & total energy\\
$B=-E=U-T$ & binding energy
\end{tabular}
\]

\section{Regular Algorithms}

In the few-body problem, close approaches of two
bodies are common, and these events require highly accurate
computations since the energies involved are large.
Thus a method that can accurately advance the motions
of few-body systems must be accurate and efficient for
the two-body problem. In addition to the classical 
KS-transformation there are some new algorithms
that satisfy this requirement. 

An additional, and most important, problem is
the roundoff error. This becomes a major problem
if e.g. center-of-mass coordinates are used and
there are close binaries and/or close encounters
in the simulated system. The cure for this problem
is the use of the chain structure, originally
applied with the KS-transformation \citep{CHAIN} 
in order to KS-regularize all the short(est) distances.
Later it became clear that the chain structure was also
beneficial in reducing significantly the roundoff error.

In this section, we first review 
the ingredients we need in the construction of the final
$N$-body algorithm. These basic algorithms produce
regular results in close approaches and  
their results can be improved to high 
precision using e.g. the \citet{BS} extrapolation method. 

\subsection{LogH, TTL and GAR}

\subsubsection{LogH}

Recently, \citet{LogH} and \citet{PretoTremaine1999} 
pointed out that the logarithmic Hamiltonian in extended 
phase space:
\begin{equation}
\LH=\ln(T+B)-\ln(U),
\end{equation}
where $B$ (the binding energy) is the momentum of time,
gives the equations of motion in the form
\begin{eqnarray}\label{timeeq}
t'&=&\doo{\LH}{B}=1/(T+B)\\  \label{coordeq}
\fat r_k'&=&\doo{T}{\fat p_k}/(T+B)\\ \label{Beq}
B'&=&\doo{U}{t}/U\\  \label{peq}
\fat p_k'&=&\doo{U}{\fat r_k}/U.
\end{eqnarray}
Here we include equation~(\ref{Beq}) although this is needed
only if there is a time-dependent potential to be added
to  the $N$-body potential $U$. 
Since the right hand sides of these equations do not 
depend on the left hand side variables, a leapfrog
algorithm is possible. That may be symbolized as
\begin{equation}
\mathbf X(h/2)\mathbf V(h)\mathbf X(h)..\mathbf V(h)\mathbf X(h/2),
\end{equation}
where $\mathbf X(s)$ means solution for the coordinate equations
(\ref{timeeq}), (\ref{coordeq}) with constant $B $ and $\fat p_k $ 
over an integration step of length$=s$:
\begin{eqnarray}
\delta t=s/(T+B);\ \ t\rightarrow t+\delta t;\ \ \fat r_k\rightarrow \fat r_k+\delta t \: \doo{T}{\fat p_k}.
\end{eqnarray}
Correspondingly $\mathbf V(s) $ signifies the operation 
\begin{eqnarray}
\widetilde{\delta t}=s/U;\ \ B\rightarrow B+\widetilde{\delta t} \: \doo{U}{t}
;\ \ \fat p_k\rightarrow \fat p_k +\widetilde{\delta t} \doo{U}{\fat r_k},
\end{eqnarray}
which solves  equations~(\ref{Beq}) and (\ref{peq}) for constant $t$ and $\fat r_k$.

For the case of two bodies only, this algorithm produces correct
trajectories, with only an $O(h^3)$ phase error. This is true
even for the collision orbit and 
the energy conservation is, in this case, typically in the machine
precision level.

\subsubsection{TTL} 

The second important ingredient in our algorithm
is the time-transformed leapfrog \citep{TTL}. 
The basic idea is to introduce a coordinate-dependent 
time transformation function $\Omega(..,\fat r_k,..)$
and a new variable $\omega$ which is supposed to have
the same numerical value as $\Omega$, but that value
is obtained from the differential equation
\begin{equation}
\dot \omega=\dot \Omega=\sum_k\doo{\Omega}{\fat r_k}\cdot { \fat v_k}.
\end{equation}
The equations of motion can be written as
\begin{eqnarray}
t'&=&1/\omega \\
\fat r_k'&=&\fat v_k/\omega\\
\fat v_k'&=&\fat A_k/\Omega\\
\omega'&=&\sum_k\doo{\Omega}{\fat r_k}\cdot { \fat v_k}/\Omega,
\end{eqnarray}
where $\fat A_k $ is the acceleration of the $k$'th particle.
The structure of these equations allows the construction of a
leapfrog algorithm:
\begin{eqnarray}\nonumber
{\rm \bf X(s)}&&\\
\delta t&=&s/\omega;\ \ t\rightarrow t+\delta t;\ \ \fat r_k\rightarrow \fat r_k+\delta t \: \fat v_k
\end{eqnarray}
\begin{eqnarray}
{ \rm \bf V(s)}&&\\
\widetilde{\delta t}&=&s/\Omega;\ 
\ \fat v_k\rightarrow \fat v_k +\widetilde{\delta t} \fat A_k;\label{vadvance}
 \ \omega\rightarrow \omega+\widetilde{\delta t} \: <\!\dot \Omega\!>,
\end{eqnarray}
where $<\dot \Omega>$ is the average of this quantity over the step,
i.e. 
\begin{equation}
<\dot \Omega>=\sum_k\doo{\Omega}{\fat r_k}\cdot(\fat v_k^{old}+\fat v_k^{new})/2.
\end{equation}
Here the superscripts ``old'' and ``new'' refer to $\fat v_k$ values
before and after the velocity advancement in the 
operation (\ref{vadvance}).

\subsubsection{GAR}
Here we concisely review the GAR method.
Following \citet{GAR}, we consider a differential equation
\begin{equation}\label{zequ}
\dot \fat z=\fat f(\fat z),
\end{equation}
and an approximation for its solution (over a short step $= h$) written as
\begin{equation}
\fat z(h)\approx \fat z(0)+\fat d(\fat z(0),h).
\end{equation}
Here the increment $\fat d(\fat z_0,h)$ can be any approximation
suitable for the particular equation in question.
If one writes, instead of (\ref{zequ}), the two equations
\begin{equation}
\dot \fat x=\fat f(\fat y);\ \ \dot \fat y=\fat f(\fat x),
\end{equation}
and solves this pair with the initial values $\fat x(0)=\fat y(0)=\fat z(0)$,
the solution obviously is $\fat x(t)=\fat y(t)=\fat z(t)$.
Using the pair of equations one can write a leapfrog as
\begin{eqnarray}\nonumber
\fat x_\half=\fat x_0+\frac{h}{2}\fat f(\fat y_0);
\: \fat y_1=\fat y_\half+h\fat f(\fat x_\half);
\: \fat x_1=\fat x_\half+\frac{h}{2}\fat f(y_1),
\end{eqnarray}
which is actually nothing but the well known modified midpoint method.
In the above one can split the advancement of $y$ in two operations to get
\begin{eqnarray}
\fat x_\half&=&\fat x_0+\frac{h}{2}\fat f(\fat y_0);
\ \fat y_\half=\fat y_0+\frac{h}{2} \fat f(\fat x_\half)\\
\fat y_1&=&\fat y_\half+\frac{h}{2}\fat f(\fat x_\half);
\ \fat x_1=\fat x_\half+\frac{h}{2}f(y_1),
\end{eqnarray}
and this can be readily generalized using the more general
increment $\fat d(\fat z,h)$. 
This results in the generalized midpoint method 
\begin{eqnarray}
\fat x_\half&=&\fat x_0+\fat d(\fat y_0,\frac{h}2);
\ \fat y_\half=\fat y_0-\fat d(\fat x_\half,-\frac{h}2),\\
\fat y_1&=&\fat y_\half+\fat d(\fat x_\half,\frac{h}2);
\ \fat x_1=\fat x_\half-\fat d(\fat y_1,-\frac{h}2).
\end{eqnarray}
This method has the great advantage that one can use any special
approximation $\fat d$ and the algorithm is time reversible
(albeit only in the extended $\fat x\:\fat y$-space) and thus suitable 
for being used as the basic integrator in an extrapolation 
method. Specifically one may stress that the increment 
$\fat d$ may be computed using LogH or TTL which produce good
approximations in case of close encounters. In addition,
appearance of velocity-dependent forces is here not
problematic, as shown by \citet{GAR}.

\section{The AR-CHAIN Algorithm}
While in the algorithms discussed above one can use any
coordinate system, the roundoff error may be a major problem
in the case of close encounters if the coordinates of the
approaching bodies are measured from a distant origin.
This problem is significantly reduced by utilizing the
chain structure. This was originally used in a KS-regularization
algorithm \citep{CHAIN} in order to get all the short distances 
regularized by the KS-transformation, but here the sole purpose is
the roundoff reduction, for which the device has proved itself.

\subsection{Basic Formulation}

Let $T={1\over 2}\sum_{k=1}^N m_k\fat v_k^2$  be the kinetic energy,
and $ U=\sum_{i<j\le N}m_im_j/r_{ij}$the potential such that
the total energy is $E=T-U$; the binding energy is $B=U-T$.

One forms a chain of particles such that the shortest relative vectors
are in the chain \citep{CHAIN}. 
We stress again that the main purpose of using the chain structure 
in this method is to reduce the (often significant) effects of 
roundoff error. 

Let us collect the chain coordinates $\fat X_k=\fat r_{i_k} -\fat r_{j_k}$  in the vector
$\fat X=(\fat X_1,\fat X_2,..,\fat X_{N-1})$ and let the corresponding velocities $\fat V_k=\fat v_{i_k}-\fat v_{j_k} $
be in the vector $ \fat V=(\fat V_1,\fat V_2,..,\fat V_{N-1})$. 
Then the Newtonian equations of motion may be formally written
\begin{eqnarray}
\dot \fat X&=&\fat V \\   \label{shortnotation}
\dot \fat V&=&\fat A(\fat X)+\fat f,
\end{eqnarray}
where $\fat A$ is the N-body acceleration and $ \fat f $ is some external 
acceleration (e.g. due to other bodies).

One may use the two equivalent time transformations \citep{GAR}
\begin{equation}
ds=[\alpha (T+B)+\beta \omega +\gamma]dt = [\alpha U+\beta \Omega +\gamma]dt,
\end{equation}
where $ s$ is the new independent variable, 
$\alpha, \beta $ and $\gamma$ are adjustable constants, $\Omega$ is an
optional function of the coordinates $\Omega=\Omega(\fat X)$, 
while the initial value
$\omega(0)=\Omega(0)$ and the differential equation
\begin{equation}
\dot \omega=\doo{\Omega}{\fat X}\cdot \fat V,
\end{equation}
determine the values of $\omega$ 
(in fact, $\omega(t)=\Omega(t)$ along the exact solution).

The time transformation thus introduced regularizes the two-body
collisions if one uses the simple leapfrog algorithm as a basic integrator;
results from which can, and must, be improved using an extrapolation method
(e.g. \citeauthor{BS} \citeyear{BS}; \citeauthor{Press} \citeyear{Press}).

It is possible to divide the equations of motion into two
categories (where derivatives with respect to the new independent
variable $s$ are denoted by a prime).

Coordinate equations:
\begin{eqnarray}
t'&=&1/(\alpha (T+B)+\beta \omega +\gamma)\\
\fat X'&=&t'\; \fat V
\end{eqnarray}

Velocity equations:
\begin{eqnarray}
\tilde t'&=&1/(\alpha U+\beta \Omega+\gamma)\\
\fat V'&=&\tilde t'\;(\fat A+\fat f)\\
\omega'&=&\tilde t'\;\doo{\Omega}{\fat X}\cdot \fat V\\
B'&=&-\tilde t'\;\doo{T}{\fat V}\cdot \fat f
\end{eqnarray}
In these equations the right hand sides do not depend on the variables
on the left. Consequently it is possible to construct a regular 
leapfrog algorithm for obtaining the solutions. 
The leapfrog results then can easily
be improved with the extrapolation method.

\subsection{Details}

\subsubsection{Finding and updating the chain}

First we find the shortest interparticle vector which 
is adopted as the first part of the chain. 
The chain is then augmented by adding the relative vector
to the particle nearest to one or the other end of 
the existing chain.
When all particles are included, they are
re-numbered along the chain as $ 1,2,..N$
for ease of programming.

To reduce roundoff problems,
the transformation from the old chain vectors 
${\bf X}_k$ to the new ones is done directly by expressing the new 
chain vectors as sums  of the old ones as in \citet{CHAIN}.

\subsubsection{Transformations}\label{transformationsection}

When the particles are renamed along the chain
as $1,2,\ldots,N$ one can evaluate 
\begin{eqnarray}
\fat X_k&=&\fat r_{k+1}-\fat r_k\\
\fat V_k&=&\fat v_{k+1}-\fat v_k.
\end{eqnarray}
The center-of-mass quantities are
\begin{eqnarray}
M&=&\sum_k m_k\\
\fat r_{cm}&=&\sum_k m_k \fat r_k/M\\
\fat v_{cm}&=&\sum_k m_k \fat v_k/M.
\end{eqnarray}

\noindent
The inverse transformation is done by simple summation
\begin{eqnarray}
\widetilde{\fat r}_1 &=&\fat 0\\
\widetilde{\fat v}_1 &=&\fat 0\\
\widetilde{\fat r}_{k+1} &=&\widetilde{\fat r}_k + \fat X_k\\
\widetilde{\fat v}_{k+1} &=&\widetilde{\fat v}_k + \fat V_k,
\end{eqnarray}
followed by reduction to the center of mass
\begin{eqnarray}
\widetilde{\fat r}_{cm}&=&\sum_k m_k \widetilde{\fat r}_k/M\\
\widetilde{\fat v}_{cm}&=&\sum_k m_k \widetilde{\fat v}_k/M\\
\fat r_{k}&=&\widetilde{\fat r}_k-\widetilde{\fat r}_{cm}\\
\fat v_{k}&=&\widetilde{\fat v}_k-\widetilde{\fat v}_{cm}.
\end{eqnarray}
Note that it is not always necessary to reduce the coordinates
to the center-of-mass system since accelerations only depend on
the differences.

\subsubsection{Equations of motion and the leapfrog}
One writes the equations of motion as
\begin{eqnarray}
\dot \fat X_k&=&\fat V_k\\
\dot \fat V_k&=&\fat F_{k+1}-\fat F_k+\fat f_{k+1}-\fat f_k,
\end{eqnarray}
where $\fat f_k$ are the individual external accelerations and
the $N$-body accelerations $\fat F_k$ are 
\begin{equation}
\fat F_k=-\sum_{j\ne k}m_j\frac{\fat r_{jk}}{|\fat r_{jk}|^3},
\end{equation}
where, for $j<k$
\begin{equation}\label{xerotus}
\fat r_{jk}=\left\{\matrix{\fat r_k-\fat r_j;\ \ \ \ \ {\rm if}\ k>j+2 \cr
                           \ \ \fat X_j;\ \ \ \ \ \ \ {\rm if}\ k=j+1\cr
			    \fat X_j+\fat X_{j+1};\ {\rm if}\ k=j+2}\right..
\end{equation}
For $k>j$ one uses $\fat r_{jk}=-\fat r_{kj}$.
In the acceleration computation the use of 
$\fat X_j$ and $  \fat X_j+\fat X_{j+1}$ reduces the roundoff
effect significantly. This is one of the most important features 
of the algorithm.

The kinetic energy is evaluated as usual
\begin{equation}
T=\frac12 \sum_k m_k \fat v_k^2,
\end{equation}
while the potential 
\begin{equation}
U=\sum_{i<j}\frac{m_im_j}{|\fat r_{ij}|},\end{equation}
is obtained along with the accelerations according to
(\ref{xerotus}).
For the time transformation function $\Omega$ it seems advantageous to use
\begin{equation}\label{Omegaequation}
\Omega=\sum_{i<j}\frac{\Omega_{ij}}{ |\fat r_{ij}|}.
\end{equation}
Here $\Omega_{ij}$ are adjustable constants (see below). 

Now one is able to evaluate the  two time transformation functions
\begin{eqnarray}
t'\ &=&\ 1/(\alpha (T+B)+\beta\omega+\gamma)\\
\tilde t'\ &=&\ 1/(\alpha U+\beta \Omega+\gamma),
\end{eqnarray}
which are equivalent along the correct solution i.e.
$t'=\tilde t'$.
The evolution of $\omega$, with $\omega(0)=\Omega(0)$, is obtained by
\begin{equation}\label{omegaequation}
\dot \omega=\sum_k\doo{\Omega}{\fat r_k}\cdot \fat v_k.
\end{equation}
In the presence of external perturbations the binding 
energy  evolves according to
\begin{equation}
\dot B=-\sum_k m_k {\fat v_k}\cdot \fat f_k.
\end{equation}
The leapfrog for the chain vectors $\fat X_k $ and $ \fat V_k$ 
can be written as the two mappings
\begin{eqnarray}\nonumber
{\bf X}(s):\ \ \ && \ \ \ \ \ \ \ \ \ \ \\
\delta t&=&s/(\alpha (T+B)+\beta \omega+\gamma)\\
t&=&t+\delta t \\
\fat X_k&\rightarrow &\fat X_k+\delta t \fat V_k\\
&& \\ \nonumber
{\bf V}(s):\ \ \ && \ \ \ \ \ \ \ \ \ \ \\
\widetilde {\delta t}&=&s/(\alpha U+\beta \Omega+\gamma)\\  \label{vupdate}
\fat V_k&\rightarrow &\fat V_k
+\widetilde{\delta t} (\fat F_{k+1}-\fat F_k+\fat f_{k+1}-\fat f_k)\\
B&\rightarrow&B+\widetilde{\delta t}\sum_k\left(-m_k<\fat v_k>\cdot\fat f_k\right)\\
\omega&\rightarrow&\omega+\widetilde{\delta t}\sum_k\doo{\Omega}{\fat r_k}\cdot <\fat v_k>,
\end{eqnarray}
where $<\fat v_k>$ is the average of the initial and 
final $\fat v$'s 
(obtained from the $\fat V$'s according to the 
equations in section (\ref{transformationsection}).
It is necessary to evaluate 
the individual velocities $\fat v_k$ because 
the expressions for $B'$ and $\omega'$ 
in terms of the chain vector velocities $\fat V_k $ are
rather cumbersome.

A leapfrog step can be written  as
\[  \fat X(h/2)\fat V(h) \fat X(h/2) \]
and a sequence of $n$ steps as
\[
\fat X(h/2)\left[\Pi_{\nu=1}^{n-1} (\fat V(h) \fat X(h))\right]  \fat V(h)\fat X(h/2).
\]
This is useful with the extrapolation method when advancing the system
over a total time interval of length $=nh$.
 
\subsection{Time Transformation Alternatives}

If one takes 
\begin{equation}
\Omega_{j}=m_im_j,
\end{equation}
then $\alpha=0,\ \beta=1,\ \gamma=0$ is mathematically 
equivalent to $\alpha=1,\ \beta=\gamma=0$ as was shown 
in \cite{TTL}.
However, numerically these are not equivalent, mainly due to
roundoff errors in updating the value of $\Omega$, and the
{\small LogH} alternative is numerically more stable. 
However, for proper treatment of small bodies
some function $\Omega$ is to be used.

For increased numerical stability in the motions of the large bodies,
and smoothing of the encounters of small bodies,
the recipe is $\alpha=1, \ \beta\neq 0 $ (but small) and
\begin{equation}
\Omega_{ij}=\left\{\matrix{={\widetilde m}^2;\ \ {\rm if }\ m_i*m_j <\epsilon\: {\widetilde m}^2 \cr
 =0;\ \ {\rm otherwise}\ \ \ \  \ \ \ \ \ \ \ }\right.,
\end{equation}
where ${\widetilde m}^2=\sum_{i<j}m_im_j/(N(N-1)/2)$ is the mean mass product  and 
$\epsilon\sim 10^{-3}$.
It may be advisable to integrate (\ref{omegaequation}) for $\omega$ 
even if $\beta=0$, in order to force the the integrator 
(extrapolation method!) 
to use short steps if $\dot \omega$ is large, thus giving higher
precision when required.
In fact,  numerical experiments suggest that in most practical cases the
parameters 
$(\alpha,\beta,\gamma)=(1,0,0)$ give the best results. Exceptions are
cases with extremely large mass ratios such that the contribution
of the small masses in the potential are negligible 
(e.g. zero masses included). 
This point, however, needs further investigation. One potential
problem in the integration of the quantity $\omega$ is that the increments
of it can be arbitrarily large if a collision of point masses occur.
In this case the roundoff errors in the value of $\omega$ are significant,
but do not affect much the results if the value of $\beta$ is small.
 One should also realize that the parameters $(\alpha,\beta,\gamma)$
 can be changed during the integration (but not during an  Bulirsch-Stoer
 integration step).

\begin{figure*}
\centerline{\includegraphics[width=0.7\textwidth]{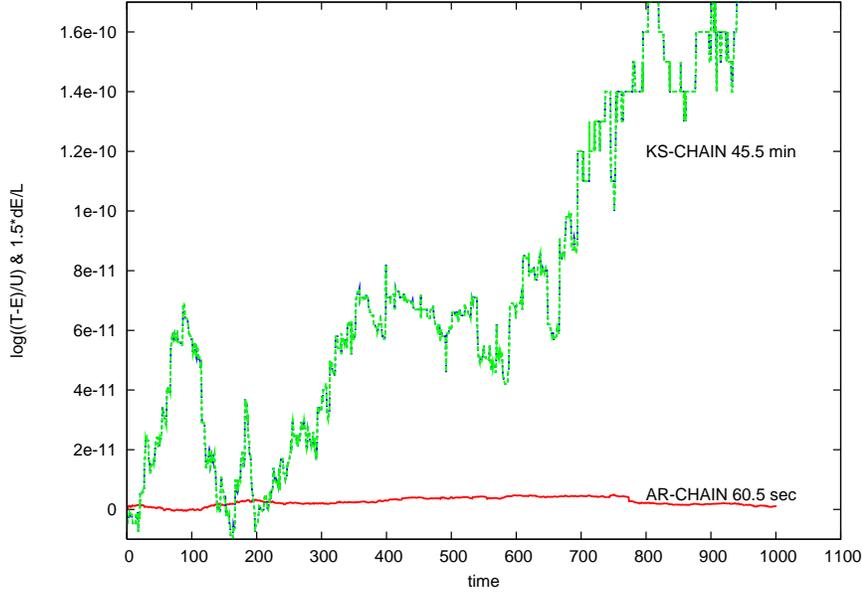}}
\caption{Evolution of $\log[(T+B)/U]$ and $1.5*dE/L$ in the integration of the initial
conditions in Table~1.
The two lines are for the two different methods: the present method (AR-CHAIN)
and the  KS-regularized chain method (KS-CHAIN). In both computations
the codes were allowed to choose automatically the stepsize.
}
\label{fig:loge}
\end{figure*}

\subsection{Velocity-Dependent Perturbations}
For the case of velocity-dependent perturbations 
$ \fat f=\fat f(\fat X,\fat V)$,
which occur e.g if one introduces relativistic post-Newtonian  terms,
related algorithms were discussed in detail by \citet{GAR} (although
mostly for the perturbed two-body problem).
Here we present those ideas in the short notation used in equation
(\ref{shortnotation}). 

\begin{description}
\item[{\bf Implicit midpoint method:}]
We have
\begin{equation}
\fat V'=\tilde t'(\fat X)\:(\fat A(\fat X)+\fat f(\fat X,\fat V)),
\end{equation}
which should be solved for constant $t,\fat X$. 
This is often not easy, but it is possible to replace the exact 
solution by the implicit midpoint method, i.e. one solves (often iteratively)
the increment of $\fat V$ from
\begin{equation}
\Delta \fat V= \widetilde {\delta t}(\fat X)(\fat A(X)
+\fat f(\fat X,\fat V_0+\half \Delta \fat V), 
\end{equation}
where $\fat V_0$ is the value of $\fat V$ before the update
$\fat V\rightarrow \fat V+\Delta \fat V$.
This operation thus replaces the one in (\ref{vupdate}),
and the mean velocities (needed in updating $B$ and $\omega$)
are obtained from $\fat V_0+\half \Delta \fat V$.

\item[{\bf Generalized midpoint method:}]

When the generalized midpoint method is used, instead of the leapfrog,
one can use a leapfrog step to obtain the increments 
$\fat d(\fat X,\fat V,s)$ and here one can use for the velocity 
the most recent available value of $\fat V$. 
One thus simply evaluates
\begin{equation}
\Delta \fat V= \widetilde {\delta t}(\fat X)\left(\fat A(X)
+\fat f(\fat X,\fat V_0)\right). 
\end{equation}
The formulation of the method then guarantees
that the final approximation has the correct symmetry (and the
correct form of the error expansion) so that the efficient Bulirsch-Stoer
extrapolation method can be used.

However, one question to be addressed here is: which of the two above
methods is most efficient?
This is problem-dependent and numerical experiments may be
necessary to answer the question. In the implementation of 
our code we start with the implicit midpoint method (which 
is most efficient if the number of iterations remains small),
and compute a long time average of the required number of iterations $Q$
recursively as
\begin{equation}
Q_{new}\rightarrow 0.99 Q_{old}+.01 Q_{now}.
\end{equation}
This slowly ``forgets'' the old values and does not grow too quickly
if there are occasional cases that require many iterations.
However, when the average number of  iterations exceeds 
some limit, then there is time to switch to the GAR that does
not require iterations (but is otherwise more expensive).
\end{description}

\section{Numerical Demonstrations}

In this section, we demonstrate the performance of the new 
algorithm and compare it with
the celebrated KS-transformed CHAIN code of \cite{CHAIN}.
We first considered a test problem consisting of one massive particle 
(the ``black hole'') with $m=1$, and seven additional particles
(``stars'') with masses $10^{-3}\le m\le 10^{-9}$.
Relativistic terms were not included.
Initial velocities were all zero so the stars moved initially on
nearly rectilinear orbits toward the black hole, but their orbits
become eccentric ellipses as they experience
perturbations from the other stars.
We set $(\alpha,\beta,\gamma) = (1,0,0)$.

We integrated the initial values given in Table~1 including 2, 3, ..,8 
bodies and carrying out integrations
up to time 1000 ($G=1$) using both the old KS-CHAIN code and the
new AR-CHAIN. Two sets of experiments were conducted,
one in which the Bulirsch-Stoer integrator automatically
chose the stepsizes (in a supposedly optimal way)
and the second with (iteration to)  
exact output times of interval $\Delta t=0.1$.

\begin{table*}
 \centering
 \begin{minipage}{140mm}
  \caption{Initial conditions for the integration of Figure 1.}
  \begin{tabular}{ccccccc}
  \hline
   $m$ & $X$ & $Y$ & $Z$ & $V_X$ & $V_Y$ & $V_Z$ \\
 \hline
$1$               & 0             & 0             & 0             &0&0&0\\
$1\times 10^{-3}$ & $-.432498862$ & $-.765730892$ & $-.432498862$ &0&0&0\\ 
$1\times 10^{-4}$ & $ .534279612$ & $.288862435$  & $.534279612$  &0&0&0\\ 
$1\times 10^{-5}$ & $-.383536991$ & $.601722629$  & $-.383536991$ &0&0&0\\ 
$1\times 10^{-6}$ & $ .233942789$ & $-.166401737$ & $.233942789$  &0&0&0\\ 
$1\times 10^{-7}$ & $ .703086026$ & $.748854732$  & $.703086026$  &0&0&0\\ 
$1\times 10^{-8}$ & $ .449061307$ & $.186538286$  & $.449061307$  &0&0&0\\ 
$1\times 10^{-9}$ & $ .320791289$ & $-.848655159$ & $.320791289$  &0&0&0\\
\hline
\end{tabular}
\end{minipage}
\end{table*}

\begin{table*}
 \centering
 \begin{minipage}{140mm}
 \caption{Automatic (free) stepsize}
  \begin{tabular}{ccc|cc}
  & AR-CHAIN & &   KS-CHAIN& \\  
  \hline
NB &  sec  & $\langle |dE/U|\rangle$   & sec  & $\langle |1.5*dE/L|\rangle$\\
2  &  0.28 & 9.0e-13&  0.6 &4.9E-13 \\
3  &  3.9  & 1.8e-13&   2.8 &4.0E-13 \\
4  &  8.1  & 2.6e-13& 8.9 &4.1E-13 \\
5  &  27.2 & 5.0e-13& 67.9 &5.8E-13 \\
6  &  43.6 & 5.5e-13&  61.8 &1.2E-12  \\
7  &  59.5 & 1.0e-12& 638.7 &2.9E-12 \\
8  &  60.5 & 3.0e-12& 2730.0& 7.8E-11\\
\hline 
\end{tabular}
\end{minipage}
\end{table*}

\begin{table*}
\centering
\begin{minipage}{140mm}
\caption{
  dt=0.1 (iteration to exact output-time)}
  \begin{tabular}{ccc|cc}
    & AR-CHAIN & &   KS-CHAIN& \\
    \hline
    NB &  sec  & $\langle|dE/U|\rangle$   & sec  & $\langle|1.5*dE/L|\rangle$\\
2 &   1.3 &  7.5E-14&  2.2& 3.0E-12\\
3  &  5.9 &  6.3E-13& 6.1& 1.9E-12\\
4  &  12.4 & 3.8E-13& 14.0& 2.4E-12 \\
5  &  32.9 & 7.6E-13& 49.6& 2.2E-12\\
6  &  38.8&  7.3E-13& 80.5& 3.1E-12 \\
7  &  56.6 & 7.4E-13& 86.8& 2.9E-12 \\
8  &  76.5&  9.4E-13& 12230.4 &1.3E-10\\
\end{tabular}
\end{minipage}
\end{table*}

The results are concisely summarized in Tables 2 and 3.
There we give for both methods the execution times 
and average values of the errors. The errors are defined
in terms of the Hamiltonians associated with the methods:
for AR-CHAIN the average was taken over $|\log((T-E)/U|\approx \delta E/U$ and for KS-CHAIN $1.5|(T-U-E)/L|\approx1.5\: \delta E/L$.
Here $L=T+U$=the Lagrangian and the factor $1.5$ is included
because in virial equilibrium $L=1.5U$. On the other
hand the system is chaotic and the solutions are, after some time,
often different (as with any method) and so the results must
be considered as giving just a general view of the
performance.  

\begin{figure}
\centerline{\includegraphics[angle=-90.,width=0.5\textwidth]{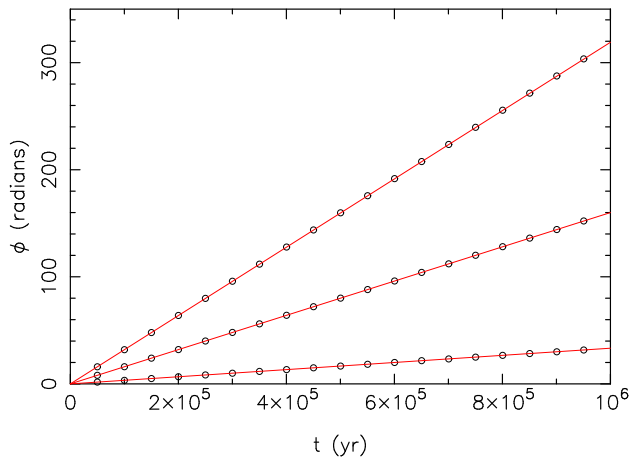}}
\caption{ Periastron advancement of a star around the Milky Way 
supermassive black hole. The semi-major axis is
$0.01$ pc and three different values of the eccentricity were tried,
 $e=(0.9,0.98,0.99)$.
Dots show the major axis orientation each 1000 orbits while
solid (red) lines show the PN2.5 prediction, equation~(\ref{eq:deltaphi}).}
\label{fig:precess}
\end{figure}

\begin{figure}
\centerline{\includegraphics[width=0.5\textwidth]{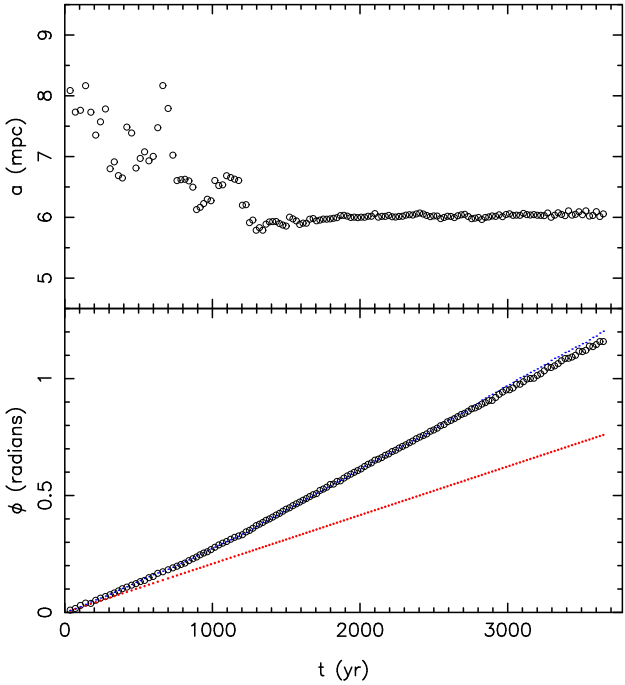}}
\caption{Integration of a star on an eccentric 
orbit around an IMBH/SMBH binary at the Galactic center.
The initial orbit elements are similar to those of the
star S0-16 \citep{Ghez:05}.
Top panel shows the semi-major axis of the star's orbit
with respect to the massive binary, lower panel shows
the advancement of the periastron.
The red (large) filled dots in this panel are the prediction
of equation~(\ref{eq:deltaphi}) assuming that $a$ and $e$ remain
fixed at their initial values.
The blue (small) dots show the predicted advancement 
if the time-dependence of $a$ and $e$ are taken into account.
This integration was continued just until the IMBH/SMBH binary 
had coalesced.
}
\label{fig:elements}
\end{figure}

\begin{figure}
\centerline{\includegraphics[width=0.5\textwidth]{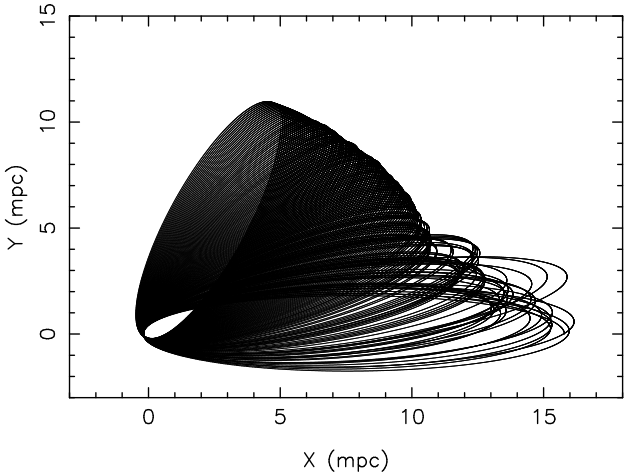}}
\caption{Configuration-space trajectory of the star whose
orbital elements are plotted in Fig.~\ref{fig:elements}.
The orbit remains in the $X-Y$ plane; the semi-major axis is 
initially parallel to the $X$ axis.
The semi-major axis changes randomly due to
perturbations from the IMBH/SMBH binary,
but after the latter has shrunk appreciably, 
$a$ remains nearly fixed and the only evolution is uniform
precession due  to the post-Newtonian terms.
}
\label{fig:orbit}
\end{figure}

One sees that the accuracies are comparable,
with some advantage in favour of the new method.
Especially we note that the new method is faster
for the case of $N=2$, i.e. a binary with eccentricity $e=1$
while the precision is essentially the same. Thus one can
conclude that this method works equally well, and perhaps
better, than the celebrated KS-transformation. This
is a consequence of the fact that for the two-body system
the logarithmic Hamiltonian + leapfrog produces an exact
trajectory with only a small error in the time, 
even for the collision orbit with eccentricity $e=1$.
 
 The accuracy of the angular momentum is typically similar
 to that of the energy. This is largely due to the fact 
 that the basic algorithm (leapfrog) used in the 
 method conserves angular momentum exactly.

However, the execution times for cases in which smaller and 
smaller bodies are included differ considerably in favour
of the new AR-CHAIN code.
 This is not very surprising, since zero masses are a singularity
for the KS-CHAIN, but AR-CHAIN can handle zero masses too.

Figure~\ref{fig:loge}  compares the energy conservation
for the two different methods
in the 8-particle integration from the initial conditions in Table 1.
The system is highly chaotic and the figure shows that 
in this very difficult case the new method is more accurate
and much faster. 

\begin{figure}
\centerline{\includegraphics[width=0.5\textwidth]{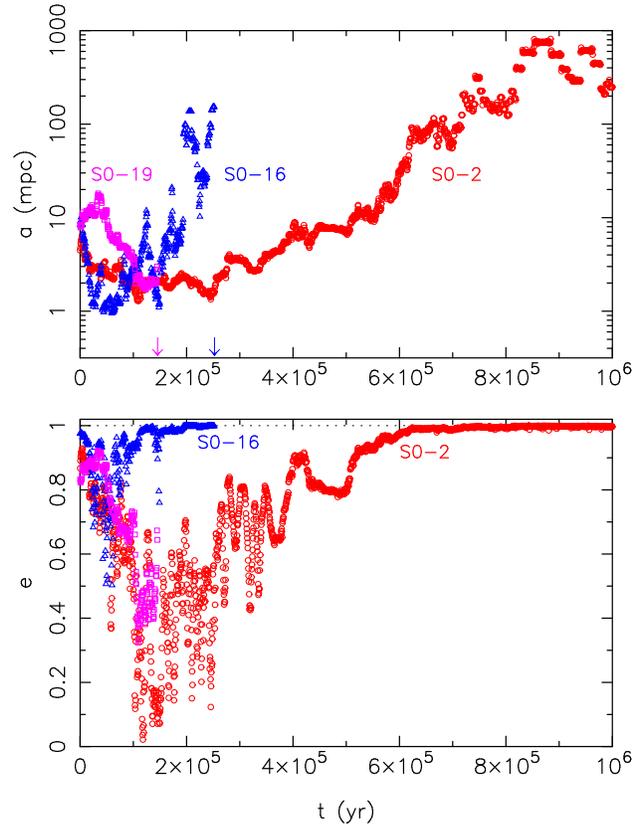}}
\caption{Evolution of the semi-major axes (top) and eccentricities
(bottom) of stars S0-2, S0-16 and S0-19 with respect to the Milky Way
supermassive black hole in a 7-particle integration that included
an intermediate-mass black hole and the five, shortest-period S0 stars.
Time zero corresponds to 2000 AD.
Arrows in the top panel indicate when S0-19 and S0-16 are ejected; 
S0-2 remained formally bound to the SMBH/IMBH binary but its 
semi-major axis gradually increased to $\sim 1$ pc.
}
\label{fig:million}
\end{figure}

Figure 2 shows a second numerical experiment that included
all post-Newtonian terms up to order PN2.5.
We considered the problem  of periastron shift of a single
star orbiting around the supermassive black hole (SMBH) 
at the center of the Milky Way.
The particle masses were $m_1=3.5\times 10^6M_\odot$ 
and $m_2=10 M_\odot=2.85\times 10^{-6}m_1$, and
the Keplerian orbit had initial semi-major axis $a=0.01$ pc, similar to
that of the ``S'' stars \citep{Ghez:05,Eisen:05}.
In units where $G=m_1=1$, and adopting $1$ mpc = $10^{-3}$ pc
as the length unit,
the speed of light is $77.19$.
Three different eccentricities were tried: $(0.9,0.98,0.99)$.
The same values of $(\alpha,\beta,\gamma)$ were used as in the
first experiment.
The integrations were continued for $10^6$ yr
or $\sim20,000$ orbital periods.
The figure plots the orientation of the Laplace-Runge-Lenz vector 
every 1000 orbits;
the solid (red) lines show the periastron advance expected
based on the PN2.5 equations,
which predict a shift each period of
\begin{equation}
\Delta \phi = {6\pi GM_1\over a(1-e^2)c^2} + 
{3(18+e^2)\pi G^2 M_1^2\over 2a^2(1-e^2)^2c^4},
\label{eq:deltaphi}
\end{equation}
or ($0.095,0.45,0.91$) degrees for $e=(0.9,0.98,0.99)$.

As a third experiment, we investigated the effect of an
intermediate-mass black hole (IMBH) on the motion of a
star orbiting around the Milky Way SMBH.
The IMBH was given a mass of $3500 M_\odot=10^{-3} M_{\rm SMBH}$,
semi-major axis $0.1$ mpc
and eccentricity $0.9$.
The large eccentricity greatly reduced the gravitational-wave
inspiral time, by a factor $\sim 10^3$ compared with a circular orbit;
inspiral required $\sim 3670$ yr or
$\sim 10^5$ initial orbital periods.
To this binary system was added a third component of mass $10M_\odot$
on an orbit having $a=8\ {\rm mpc}\approx 1600$ AU and 
$e=0.974$ with respect to the center of mass of the SMBH/IMBH binary.
These orbital elements are similar to those
inferred for the star S0-16  \citep{Ghez:05}.
The star and IMBH were coplanar with aligned angular momenta.
The star completed $\sim 10^2$ orbits during the time of IMBH inspiral.
Figure~\ref{fig:elements} shows the evolution of the star's 
semi-major axis and its orbital orientation during this time,
and the configuration-space trajectory is plotted in Figure~\ref{fig:orbit}.
Initially, the apoastron distance of the IMBH 
is $\sim 0.19$ mpc, similar to the periastron distance
of the star, $\sim 0.21$ mpc, so that close interactions are allowed.
The primary influence was found to be on the star's semi-major axis;
the eccentricity of the star's orbit changed only slightly.
The star's periastron advancement was found to be well predicted
by equation~(\ref{eq:deltaphi}) if the time dependence of $a$ and $e$ 
were taken into account (Fig.~\ref{fig:elements}).
A second integration without the PN terms confirmed that 
the IMBH itself contributes only slightly to the star's 
precession rate for this configuration.

Our final numerical experiment was a one-million-year
integration of a seven-body system consisting of the MW SMBH;
an IMBH of mass $10^{-3} M_{\rm SMBH}$; and the five, shortest-period
S stars: S0-1, S0-2, S0-16, S0-19, S0-20 \citep{Ghez:05}.
Stellar masses were set to $15 M_\odot$ and the
initial positions and velocities were determined at year 2000 AD
using the Keplerian orbital elements given in Table 3 of Ghez et al. (2005).
The IMBH orbit was assigned an initial eccentricity of $0.9$
and semi-major axis of $1$ mpc, compared with $4\ {\rm mpc} \lesssim a
\lesssim 25\ {\rm mpc}$ for the stars.
Post-Newtonian terms were included, causing the orbit of the
IMBH to precess rapidly and to fill the annulus 
$0.1\ {\rm mpc}\lesssim r \lesssim 1.9\ {\rm mpc}$; 
for this choice of $(a,e,M_{\rm IMBH})$ the gravitational
wave inspiral time is $\sim 10^8$ yr, much greater than the
length of the integration.
Three of the included stars, S0-2, S0-16, S0-19, have periastron
distances that intersect the IMBH's orbital annulus and so each of these stars
was able to interact closely with the IMBH, although many orbital periods
were required before close encounters occurred.
S0-19 was the first to achieve positive energy,
at $t\approx 147,500$ AD; before ejection the star's orbit
evolved toward small $a$ and $e$.
S0-16 was the next to escape, at $t\approx 254,500$AD;
this star moved into a highly eccentric orbit before being ejected.
S0-2 remained bound to the SMBH/IMBH  binary but its semi-major axis  increased
gradually to $\sim 1$ pc, roughly equal to the radius of influence
of the SMBH
(if the latter were embedded in the Galactic bulge),
so in a practical sense it too may be considered to have escaped.
Figure~\ref{fig:million} shows the evolution of $a$ and $e$ for
the three stars.
Experiments like these could be used to constrain the mass
and orbital parameters of a putative IMBH near the Galactic
center.

\section{Concluding Remarks}

We have demonstrated the logH method
(with leapfrog and Bulirsch-Stoer-extrapolation)
is at least equally good, and for some systems better, 
than the KS-CHAIN method. 
The optional TTL (which we use only to provide some regularization for the
 occasional close encounters of very small bodies) seems to have some 
 problems that may have to do with roundoff error when the function
 $\omega$ first increases then decreases by a large amount in case
 of very close encounters.
However, we point out that this round-off problem is reduced almost to
non-existence by including
in the TTL-function $\Omega$ only the interactions between small bodies
 and using only a small factor for this quantity in the time
 transformation function. Since the TTL function derivative is
 explicitly integrated it automatically reduces the stepsize
 in case of close approach of small bodies thus providing more careful
 integration of such events even if the coefficient $\beta$ is negligible.
 This is due to the properties (sensitivity) of the
 Bulirsch-Stoer extrapolation method that sees any tiny irregularity
 in the data and modifies the stepsize in response. This property
 of the BS-extrapolation method also helps if the leapfrog  algorithm
 happens to evaluate the derivatives too close to collision of some pair of
 bodies: the step is rejected and re-evaluated with a different stepsize.
 This procedure normally avoids the repetition of a such a numerical accident. 
 Finally, our code gives the option of not using time transformation at all. 
 Even in this alternative the present code is much better than the straightforward use of
 center-of-mass coordinates. 
 The reason is that the chain coordinates  reduce
 the roundoff significantly. 
 This is one of the great advantages of the chain structure.

\bigskip\bigskip\bigskip

\acknowledgments
DM was supported by grants AST-0420920 and AST-0437519 from the NSF
and grant NNX07AH15G from NASA.

\end{document}